\documentclass[12pt]{article}
\usepackage{jasa_macro}

\usepackage{microtype}

\usepackage{color,url}
\usepackage{algorithm}
\usepackage{algpseudocode}
\usepackage{natbib}

\usepackage{epsfig}
\usepackage{epstopdf}

\begin{document}

\title{Bayesian Inference of Selection in the Wright-Fisher Diffusion Model}

\author{Jeffrey J. Gory \quad Radu Herbei \quad Laura S. Kubatko \\
The Ohio State University}

\date{}

\maketitle

\newpage
\begin{center}
\textbf{Abstract}
\end{center}
The increasing availability of population-level allele frequency data across one or more related populations necessitates the development of methods that can efficiently estimate population genetics parameters, such as the strength of selection acting on the population(s), from such data. Existing methods for this problem in the setting of the Wright-Fisher diffusion model are primarily likelihood-based, and rely on numerical approximation for likelihood computation and on bootstrapping for assessment of variability in the resulting estimates, requiring extensive computation. Recent work \citep{jenkinsspano2015} has provided a method for obtaining exact samples from general Wright-Fisher diffusion processes, enabling the development of methods for Bayesian estimation in this setting.  We develop and implement a Bayesian method for estimating the strength of selection based on the Wright-Fisher diffusion for data sampled at a single time point.  The method utilizes the work of \citet{jenkinsspano2015} to develop a Markov chain Monte Carlo algorithm to draw samples from the joint posterior distribution of the selection coefficient and the allele frequencies. We demonstrate that when assumptions about the initial allele frequencies are accurate the method performs well for both simulated data and for an empirical data set on hypoxia in flies \citep{zhouetal2011}, where we find evidence for strong positive selection in a region of chromosome 2L previously identified by \citet{ronenetal2013}.  We discuss possible extensions of our method to the more general settings commonly encountered in practice, highlighting the advantages of Bayesian approaches to inference in this setting.
\vspace*{.3in}


\newpage
\section{Introduction}

The Wright-Fisher model \citep{fisher1930,wright1931} is the most widely used model for the evolution of genetic variation at a locus over time. In its most basic form, the Wright-Fisher model assumes a population of $N$ individuals, each with two possible genetic variants (called alleles) at the locus of interest, for a total of $2N$ alleles in the population.  Of those $2N$ alleles,  $X$ are assumed to be of type $A$, and the remaining $2N-X$ alleles are assumed to be of type $a$. The Wright-Fisher model traces the proportion of $A$ alleles across generations under the assumption of discrete, non-overlapping generations of constant size, in which the next generation of $2N$ alleles is produced by sampling from the current  generation at random with replacement.  Under this model, it is easy to see that the number of $A$ alleles at time $t$, denoted $X_t$, is a discrete-time Markov process with transition probabilities given by the Binomial distribution with $2N$ trials and probability  $q_{t-1} = X_{t-1}/(2N)$ of success.  In this model, states $0$ and $2N$ are absorbing states.

Application of the basic Wright-Fisher model is limited, however, in several ways. For example, it deals only with variation at a single locus in a single population, and it neglects other important aspects of the evolutionary process, such as selection for advantageous or against deleterious mutations, changes in population size over time, and the generation of new allelic variants through spontaneous mutation. While these processes may be straight-forwardly modeled, inference in the discrete-time framework in these more complex situations is generally not computationally tractable. 

For this reason, two common continuous-time approximations are typically employed. The first is generally referred to as Kingman's coalescent process \citep{kingman1982a, kingman1982b,kingman1982c}, and arises by examining the limiting distribution of the time at which a collection of sampled alleles last shared a common ancestor as the population size becomes large relative to the sample size. The coalescent approach has proven extremely useful for species-level phylogenetic inference for moderate numbers of species or populations (generally, three to 20 species/populations) under the basic Wright-Fisher model in which the effects of selection are negligible and in which no gene flow between distinct species occurs \citep{edwards2009,knowleskubatko2010}.  However, inference in the presence of selection and/or migration between species or populations is much more difficult using Kingman's coalescent  \citep{wakeleyhey1997,nielsenwakeley2001,heynielsen2007,wakeley2009}, and both simulation and empirical studies indicate that the inferences obtained under models that ignore these processes can be biased in important ways \citep{eckertcarstens2008,leacheetal2014,burbrinkguiher2015}.

The second continuous-time approximation to the Wright-Fisher model is based on modeling $Q_t$, the proportion of $A$ alleles in a population of size $N$ at time $t$,  as a diffusion process. This process is defined through the following stochastic differential equation (SDE) 
\begin{equation} 
\label{eq:WFgeneric}
\left\{
\begin{array}{l}
dQ_t = \beta(Q_t)dt + \sqrt{Q_t(1-Q_t)} dB_t\\
Q_0 = q_0, \quad 0\le t\le T
\end{array}
\right.
\end{equation}
which will be referred to as the Wright-Fisher (WF) diffusion. In \eqref{eq:WFgeneric}, the initial condition is $q_0\in(0,1)$, $\{B_t, t\ge 0\}$ is a standard Brownian motion process and the coefficient $\beta: [0,1] \rightarrow \mathbb{R}$ is assumed to satisfy sufficient smoothness conditions such that \eqref{eq:WFgeneric} admits a unique weak solution. Such conditions can be found in Section 5.3 of \citet{karatzas2012brownian}. In applications, the drift function $\beta(\cdot)$ is selected to incorporate complex evolutionary processes such as selection and mutation.   Because such processes are easily specified in this approach, the WF diffusion has been increasingly applied to settings in which the number of populations is small (typically on the order of one to three populations) and primary interest is on inferring the contribution of these evolutionary processes to the current genetic configuration of the population(s).  

Practical use of (\ref{eq:WFgeneric}) is complicated by the fact that the transition density for the WF diffusion does not have a closed form.  Current approaches using this model are therefore based on approximations of this density.  A typical solution is to approximate the transition density using a numerical scheme for the SDE \citep[see][for a classical overview]{kloedenplaten}.  Other recent approaches to estimating the transition density of an SDE include methods based on importance sampling \citep[see][and the references therein for a review]{Pedersen1995, elerian2001likelihood, brandt2002simulated, DurhamGallant2002, lin2010generating}, closed-form approximations based on a Hermite polynomials expansion \citep{AitSahalia,ait2008closed}, a strategy based on approximating the solution of the Kolmogorov forward equation \citep{lo1988maximum}, and exact sampling \citep{EAEstimate}.  Another approach, specific to approximating the transition density of the WF diffusion, involves truncation of a spectral representation of the target density \citep{songetal2012,steinruckenetal2013}. 

Given the wealth of genetic data currently available as a result of ever-improving sequencing technologies, an additional challenge for the diffusion-based inferential framework is the desire to use data from many loci simultaneously for the inference of evolutionary process parameters. This is typically done by making the assumption that the loci included in the study are unlinked, and thus independent, as would typically be the case for single nucleotide polymorphism (SNP) data collected across the genome.   Under this assumption, the likelihood  is computed by multiplying single-locus likelihoods computed under the model across loci.  When this assumption is violated and loci are actually linked, the likelihood is viewed as a composite likelihood, and inference proceeds as usual.

This framework has been considered in a variety of contexts. \citet{williamsonetal2005} considered specification of the drift coefficient to include the effect of selection, and fit models that allowed for changes in population size throughout time.  To carry out the computations required by the WF diffusion, the Crank-Nicolson approximation was used to solve the relevant SDE in order to compute the likelihood. Estimation of the selection coefficient and population size parameters was then carried out in a likelihood framework. These authors later extended the method to include multiple populations with the possibility of gene flow between populations, again using a finite differencing method to provide a numerical solution to the SDE \citep{gutenkunstetal2009}.  As before, inference was carried out in a likelihood framework, with variance estimates and significance values for hypothesis testing obtained using bootstrapping.  These methods were implemented in the program $\partial a \partial i$ for up to three populations \citep{gutenkunstetal2010}, and  have been applied to a variety of organisms, including humans, cattle, rice, and bees \citep{robinsonetal2014}.  In addition, recent attention has been given to methodology for efficient computation of estimates of uncertainty in parameter estimates, with applications to model selection \citep{coffmanetal2015, robinsonetal2014}.

Despite a great deal of previous work in this area, inference in a Bayesian framework has been limited.  \citet{schraiberetal2016} and \citet{ferreradmetllaetal2016} applied Bayesian techniques to infer selection in allele frequency time series data.  Their methods require approximation of the transition density of the WF diffusion.  \citet{jenkinsspano2015} recently developed an algorithm for obtaining exact draws from the WF diffusion with general drift function, using draws from the neutral WF diffusion with mutation.  The work of \citet{jenkinsspano2015} thus enables the development of Markov chain Monte Carlo (MCMC) approaches to inference of evolutionary process parameters from genome-wide SNP data based on the composite likelihood, but without the need for numerical approximation of the solution to the SDE.

Here, we implement one such method for a single-population model in which interest is focused on estimation of the selection parameter.  Unlike \citet{schraiberetal2016}, \citet{ferreradmetllaetal2016}, \citet{folletal2015}, and \citet{malaspinasetal2012}, who considered time series data, we assume that allele frequency data are only available at a single time point.  We note that previous work in this single time point setting has focused primarily on likelihood- or summary statistic-based inference, rather than on Bayesian approaches. In particular, \citet{nielsenetal2005} proposed a composite likelihood statistic for the detection of selective sweeps that allowed estimation of the magnitude of the selection coefficient as well as the location of the sweep.  This method, called SweepFinder, was later made more computationally efficient in an implementation called SweeD \citep{pavlidisetal2013}.  Similarly, the linkage disequilibrium-based method of \citet{kimnielsen2004} has recently been implemented with improved computational efficiency in the software OmegaPlus \citep{alachiotisetal2012}.  Finally, the methodology proposed in \citet{zivkovicetal2015} could also be applied to data sampled at a single time point.  All of these approaches, however, involve computation of statistics (e.g., composite likelihoods) at various positions in the genome, with no naturally associated measure of variance, aside from those obtained via bootstrap procedures, as in \citet{zivkovicetal2015}.  We thus propose a Bayesian approach to the problem of data sampled at a single time point. 

In the next section, we describe the details of our proposed model, as well as the computational approach we take to implement inference in an MCMC-based Bayesian framework.   We apply the method to simulated data to study its performance, which provides insight into the role of the prior distribution in inference under the model. We then apply the method to an empirical data set \citep{zhouetal2011} consisting of SNP data on flies that were subjected to either a hypoxic environment or a control (e.g., normal oxygen) environment for 200 generations, with the goal of estimating the extent of selection for the flies that were oxygen-deprived. We compare our results with those of \citet{ronenetal2013}, who analyzed the same data. Finally, we discuss potential extensions of our results to more general settings, as well as the advantages of inference in the Bayesian setting for these problems.

\section{Methods}
\label{sec:approach}

Suppose that a total of $n$ individuals are sampled, and that the type of allele ($A$ or $a$) is recorded for each individual at each of $K$ loci.  At locus $k$, the number of individuals with allele $A$ is denoted $y_k$.  Let ${\mathbf y} = (y_1, y_2, \ldots , y_K)$ be the complete vector of data across all $K$ loci. For each locus, $y_k$ is assumed to be an observation from a Binomial distribution with $n$ trials and probability $q_k$ of success.  Note that $q_k = q_{T,k}$ represents the proportion of $A$ alleles in the population at locus $k$ at time $T$ -- the time at which the sample is collected -- however, we will suppress the dependence on $T$ throughout the manuscript, in order to simplify notation.  Conditional on the vector $\mathbf{q} = (q_1, q_2, \ldots , q_K)$, the likelihood is given by 
\begin{equation} \label{eq:lik}
g(\mathbf{y} | \mathbf{q}) = \prod_{k=1}^{K} \binom{n}{y_k} q_k^{y_k} (1-q_k)^{n-y_k} .
\end{equation}
Note that this assumes conditional independence of the data across loci; if the loci are not independent given the allele frequencies, then the likelihood in (\ref{eq:lik}) can be viewed as a pseudo-likelihood. This is the same likelihood function used in other applications \citep[e.g.,][]{williamsonetal2005}.

The allele frequencies $q_k$, $k=1,\dots, K$ are assumed to be realizations (at time $T$) from WF diffusion processes with  starting values $q_{0,k}$,\ $k=1,\dots, K$. Often, $T$ is a parameter of interest in the model, however, in this case we will assume that $T$ is fixed and known.  The drift coefficient $\beta(\cdot)$ is specified to incorporate the evolutionary processes of interest and associated model parameters.   The goal of the inference procedure is to estimate the evolutionary model parameters given the data $\mathbf{y}$ and carry out standard inference (e.g., hypothesis testing).

Here we consider a model in which selection and mutation operate within a single population.  Selection refers to the evolutionary mechanism in which one allele is preferred over other potential alleles to survive to the next generation, while mutation refers to the origination of new types of alleles within the population.  Over time, both selection and mutation can act to alter the frequency of allele $A$ at a particular locus in the population beyond what is expected under the basic Wright-Fisher model that does not include these processes.   The WF diffusion corresponding to the model with mutation and selection is given by 
\begin{equation}
\left\{
\begin{array}{l}
\label{eq:WFmutsel}
dQ_t = [sQ_t(1-Q_t) + \frac{1}{2}(\theta(1-2Q_t))]  dt \\
\hspace{2.9cm}  + \sqrt{Q_t(1-Q_t)} dB_t,\\
Q_0 = q_0, \quad 0\le t\le T, 
\end{array}
\right.
\end{equation}
where $s$ is the population-scaled selection parameter and $\theta$ is the population-scaled mutation parameter \citep[see][Chapter 5]{etheridge2011some}.  For each locus $k$, define the density of the allele frequency at time $T$ under this WF diffusion by $f(q_k | s, \theta, q_{0,k}, T)$, where $q_{0,k}$ is the initial allele frequency at locus $k$.

Assuming that $\theta$, $T$, and $\mathbf{q}_0 = (q_{0,1}, q_{0,2}, \ldots , q_{0,K})$ are known, we focus on estimating the extent of selection via inference of the parameter $s$. We take a Bayesian viewpoint and base our statistical inference on the posterior density function
\begin{equation}
\label{eq:posterior}
\pi(s, \mathbf{q} \giv \mathbf{y}) \propto g(\mathbf{y}\giv\mathbf{q}){\mathbf f}(\mathbf{q}\giv s, \theta,\mathbf{q}_0,T)\pi_s(s),
\end{equation}
where we use the notation
$$
{\mathbf f}(\mathbf{q}\giv s, \theta,\mathbf{q}_0,T) = \prod_{k=1}^K f(q_k | s, \theta, q_{0,k}, T)
$$
and $\pi_s(s)$ is a specified prior distribution on the selection parameter $s$. Subsequently, we use an MCMC algorithm to explore the distribution given by \eqref{eq:posterior}.

To that end, note that a typical Metropolis within Gibbs (MwG) sampler would alternate between updating $s$ and $\bfq$, conditional on everything else, using some proposal-acceptance mechanism. However, a direct implementation of this procedure suffers from several issues. For example, during the ``update $\mathbf q$ given $s$'' step, under the assumption that the proposal distribution for $\bfq$ is symmetric, the MwG accept-reject mechanism requires evaluating the acceptance ratio
\begin{equation*}
\alpha = \frac{g(\bfy\giv\bfq'){\mathbf f}(\bfq'\giv s,\theta,\bfq_0,T)}{g(\bfy\giv\bfq_n){\mathbf f}(\bfq_n\giv s,\theta,\bfq_0,T)},
\end{equation*}
where $\bfq_n$ and $\bfq'$ are the current and proposed states respectively. However, since the transition density for the WF diffusion $f(q_k\giv s,\theta,q_{0,k},T)$ does not have a closed form, the ratio above cannot be evaluated exactly in practice.

Most of the methods described in Section~1 for estimating $f(q_k\giv s,\theta,q_{0,k},T)$ come with a significant computational cost.  Such methods are typically designed to work in a maximum likelihood estimation setting, where one would have a relatively small number of transition density evaluations. In a Bayesian setting however, simulating a Markov chain would require a very large number of transition density evaluations and thus many of these methods become impracticable.  

In this manuscript we suggest approaches based on exact sampling. Our goal is to use the transition density $f(\cdot)$ in an MCMC setting and thus we require computational efficiency. We propose two sampling strategies: (i) an exact sampler which avoids evaluating the transition density $f(\cdot)$ entirely but suffers from some computational overhead and (ii) an efficient approximate sampler based on exact simulations of the diffusion process \eqref{eq:WFmutsel}. In both cases, our method relies on the ability to obtain exact draws from the density $f(\cdot \giv s,\theta,\bfq_0,T)$. We explain how this is done in detail in the Appendix. Henceforth, we assume that we can simulate variates
$$
{\mathsf Q}_k\sim f(\cdot \giv s,\theta,q_{0,k},T) \quad \mbox{for} \quad k=1,\dots, K.
$$
Although such simulation is possible, it can be extremely time-consuming.

\vspace{0.5cm}
\noindent
\textbf{Exact Sampler}. Our exact MCMC sampler uses a joint proposal mechanism as follows. Let $(s_n, \bfq_n)$ be the current state of the Markov chain exploring the distribution \eqref{eq:posterior} and consider simulating a new proposed state $(s', \bfq')$ from the (joint) prior distribution as follows:
\begin{algorithm}
\caption{Joint simulation of $(s',\bfq')$ from the prior model}
\label{alg:prior}
\begin{algorithmic}[1]
\State Draw ${\mathsf S}'\sim \pi_s(\cdot)$;
\State Given ${\mathsf S}' = s'$, draw ${\mathsf Q}'_{k} \sim f(\cdot\giv s', \theta,q_{0,k},T)$ for $k=1,\dots,K$;
\State Set $\bfq'=({\mathsf Q}_{1}', \dots {\mathsf Q}_{K}')$;
\State \textbf{Return} $(s', \bfq')$.
\end{algorithmic}
\end{algorithm}

In this case one can derive the acceptance ratio for $(s', \bfq')$ to be
$$
\alpha = \frac{g(\bfy\giv\bfq')}{g(\bfy\giv\bfq_n)},
$$
which does not depend on $f(\cdot \giv s,\theta,q_{0,k},T)$ and can therefore be easily evaluated. Exact draws from $f(\cdot \giv s,\theta,q_{0,k},T)$ are used to generate proposals, but it is not necessary to evaluate $f(\cdot \giv s,\theta,q_{0,k},T)$ to compute the acceptance ratio.  Thus, unlike existing methods, this strategy does not require one to approximate the transition density.  

There are, however, a number of drawbacks that one should consider. Given that this algorithm is essentially an independent Metropolis-Hastings sampler, it is well known that it may mix very poorly. Further, the second sampling step requires one to simulate the SDE \eqref{eq:WFmutsel}. As mentioned above and discussed in greater detail in Section~\ref{sec:results}, such simulation can be extremely time-consuming.  Since each update in the MCMC procedure requires this sampling step, simulation of the SDE becomes a bottleneck in the algorithm. Due to the combination of poor mixing and a slow sampling step, we found this algorithm to be impracticable computationally, especially when studying a large number of loci simultaneously. Nonetheless, given sufficient computing resources, we believe this procedure would perform adequately.

\vspace{0.5cm}
\noindent
\textbf{Approximate MwG}. A natural solution to the issues described above is to find an analytical expression for the transition density $f(\cdot\giv s,\theta,\mathbf{q}_0,T)$,  which would allow one to implement a standard MwG algorithm.  As discussed earlier, this is a difficult task.  To take advantage of our ability to sample exactly from $f(\cdot\giv s,\theta,\mathbf{q}_0,T)$, we suggest approximating the transition density using a kernel density estimate   
\begin{equation}
\label{eq:gaussiankde}
\tilde{f}(q \giv s,\theta,q_{0,k},T)=\frac{1}{M} \sum_{i=1}^M \frac{1}{h}\phi\bigg{(}\frac{q-{\mathsf Q_i}}{h}\bigg{)},
\end{equation}
where $\phi(\cdot)$ is the density of the standard Gaussian distribution, $h$ is a bandwidth parameter, and ${\mathsf Q}_1, \dots, {\mathsf Q}_M$ are independent and identically distributed draws from the density $f(\cdot\giv s,\theta,q_{0,k},T)$.  We choose $h\approx M^{-1/5}$ according to Scott's Rule \citep{scott1992} and in our simulations we use $M = 10,000$, see Section \ref{sec:results}. It is known that the mean integrated squared error in this case is 
$$E\left( \int_0^1(f(q) - \tilde f(q))^2\ dq \right)  = {\mathcal O}(M^{-4/5})\ ,$$
where the expected value is taken with respect to the Monte Carlo draws. Once the transition density has been approximated, the posterior distribution \eqref{eq:posterior} can be replaced by
\begin{equation}
\label{eq:post}
\tilde \pi(s, \mathbf{q} \giv \mathbf{y}) \propto g(\mathbf{y}\giv\mathbf{q})\tilde {\mathbf f}(\mathbf{q}\giv s, \theta,\mathbf{q}_0,T)\pi_s(s),
\end{equation}
where
\begin{equation}
\label{eq:densityest}
\tilde{\mathbf f}(\mathbf{q}\giv s, \theta,\mathbf{q}_0,T) = \prod_{k=1}^K \tilde f(q_k | s, \theta, q_{0,k}, T)\ .
\end{equation}
The distribution \eqref{eq:post} can then be explored directly using a standard MwG algorithm.  

This approximate MwG algorithm requires a kernel density estimate of $f(q_k \giv s,\theta,q_{0,k},T)$ for each potential set of proposed values for $s$, $\theta$, $q_{0,k}$, and $T$.  For this to be feasible we must restrict ourselves to a finite set of plausible values for these parameters.  We choose to treat both $\theta$ and $T$ as fixed and known, and to assume that $s$ and $q_{0,k}$ are contained in the finite sets $\mathcal{S}$ and $\mathcal{Q}_0$, respectively.  The initial allele frequencies are generally not known in practice, so it is not reasonable to treat $\mathbf{q}_0$ as fixed and known.  Instead, we view each $q_{0,k}$ as an independent draw from a discrete distribution $\pi_q(\cdot)$ that puts positive mass on $\mathcal{Q}_0$.  This allows us to account for uncertainty in $\mathbf{q}_0$, but presents a challenge in that we cannot estimate $s$ and $\mathbf{q}_0$ simultaneously.  We therefore ``integrate out" $q_{0,k}$ to leave $s$ as the only parameter to be estimated.  

Ultimately, for each $s \in \mathcal{S}$ and $q_{0,k} \in \mathcal{Q}_0$, we use (\ref{eq:gaussiankde}) to produce a kernel density estimate $\tilde{f}(q \giv s,\theta,q_{0,k},T)$ of the transition density and then sum over $\mathcal{Q}_0$ to compute an alternative estimate $\hat{f}(q \giv s,\theta,T)$ that does not depend on $q_{0,k}$.  Specifically, we compute 
\begin{equation} \label{eq:sumoverq0k}
\hat{f}(q_k|s,\theta,T)=\sum_{q_{0,k} \in \mathcal{Q}_0} \tilde{f}(q_k \giv s,\theta,q_{0,k},T) \pi_q(q_{0,k}).
\end{equation} 
The density estimate $\hat{f}(q_k \giv s,\theta,T)$ replaces $\tilde{f}(q_k \giv s,\theta,q_{0,k},T)$ in (\ref{eq:densityest}) and the algorithm proceeds as if $\mathbf{q}_0$ were known.

The upfront computational cost of this approach is substantial as one must simulate many draws from many transition densities in order to generate a sufficient number of accurate kernel density estimates with which to effectively explore the parameter space.  However, once these density estimates have been produced, the MCMC algorithm is able to run efficiently.  This is in contrast to the exact sampler, which does not have the same upfront costs but performs each update inefficiently.  Further, by updating the parameters individually, conditional on all other parameters in the model, the MwG sampler mixes better than the exact sampler and therefore requires far fewer updates to obtain a sample of approximately independent draws from the target posterior.

\section{Results}
\label{sec:results}

\noindent
\textbf{Simulated Data.} In order to assess the viability of our approach we first applied it to simulated data with known selection parameter.  Specifically, for each of three different selection parameters ($s=0.0$, $s=5.5$, and $s=11.0$) we simulated 500 data sets with $n=200$ individuals and $K=324$ loci.  This was accomplished by using the algorithm described in the Appendix to simulate each $q_k$ from the SDE given in (\ref{eq:WFmutsel}) and drawing each corresponding $y_k$ from the appropriate Binomial distribution.  Model parameters were given values matching those used with the empirical fly data in the example described below.  Specifically, for each data set we set $\theta=0.00014$ and $T=0.1$, and let $\mathbf{q}_0$ follow the ending distribution of allele frequencies for the neutral fly population in the example. 

We conducted two sets of simulations.  In both cases we defined the set of plausible values for $s$ as $\mathcal{S}=\{-12.0,-11.5,\dots,17.0\}$ and assumed a distribution $\pi_q(\cdot)$ for the initial allele frequencies that put mass on the set $\mathcal{Q}_0=\{0.01,0.02,\dots,0.99\}$.  As such, we computed $59 \times 99 = 5,841$ kernel density estimates $\tilde{f}(q_k \giv s,\theta,q_{0,k},T)$ with one corresponding to each combination of $s \in \mathcal{S}$ and $q_{0,k} \in \mathcal{Q}_0$.  Each of these density estimates was based on 10,000 draws from the true transition density, produced via exact simulation of the SDE in (\ref{eq:WFmutsel}). We then summed over $\mathcal{Q}_0$ as in (\ref{eq:sumoverq0k}) to obtain an estimate $\hat{f}(q_k \giv s,\theta,T)$ for each of the $59$ values of $s$ in $\mathcal{S}$.  

Generating the samples required to compute these density estimates took nearly 1,600 hours on a NVIDIA Tesla C2050 GPU computing card.  However, a disproportionate amount of this time was required to generate samples for the largest values of $s$.  It took about 607 hours for $s=17.0$ alone, about 384 hours for $s=16.5$, and about 215 hours for $s=16.0$.  In contrast, generating the required samples for all values of $s$ in the set $\{-8.0,-7.5,\dots,8.0\}$ took a total of just 32 minutes.  Large values of $s$ require more time because the proposals in the acceptance-rejection algorithm are generated from a process with $s=0.0$.  Thus, as $s$ increases in magnitude there is a bigger discrepancy between the process used for proposals and the target process, so far fewer proposals are accepted.   

For the first set of simulations we defined $\pi_q(\cdot)$ as a discretized version of the distribution of $\mathbf{q}_0$ that was used to generate the data whereas in the second set of simulations we defined $\pi_q(\cdot)$ as a discrete Uniform distribution over the set $\mathcal{Q}_0$.  Altering the assumed distribution for $\mathbf{q}_0$ allows us to assess the importance of this choice and its impact on inference for $s$.

For all simulations we placed a discrete Uniform prior on $s$ so that $\pi_s(s)=\frac{1}{59}$ for each $s \in \mathcal{S}$.  Our proposal density for $s$ was a discrete Uniform centered on the current value of $s$ that put positive mass on the five grid values on either side of the current value \big{(}$p(s|s_n)=\frac{1}{11}$ for $s \in \{s_n-2.5,s_n-2.0,\dots,s_n+2.5\}$ and $p(s|s_n)=0$ otherwise\big{)}. Proposals for each $q_k$ came from a $\mathrm{N}(q_{n,k},\sigma_q^2)$ distribution where $q_{n,k}$ was the current value for $q_k$ and $\sigma_q^2=0.05^2$ was the proposal variance.  Since these proposal densities are symmetric they do not play a role in the acceptance ratios of the MwG algorithm.  For each simulated data set, a Markov chain was run for 10,000 steps.  The first 1,000 steps were discarded as burn-in and every tenth step thereafter was retained to form a sample of size 900 from the posterior distribution.  Trace plots and autocorrelation plots for a random subsample of the replicates (not shown) suggest that the Markov chains converge to a stationary distribution within 1,000 steps and that retaining only every tenth step eliminates virtually all of the autocorrelation among values in the posterior sample.  

\begin{figure}[!h]
\begin{center}
\includegraphics[width=3.2in]{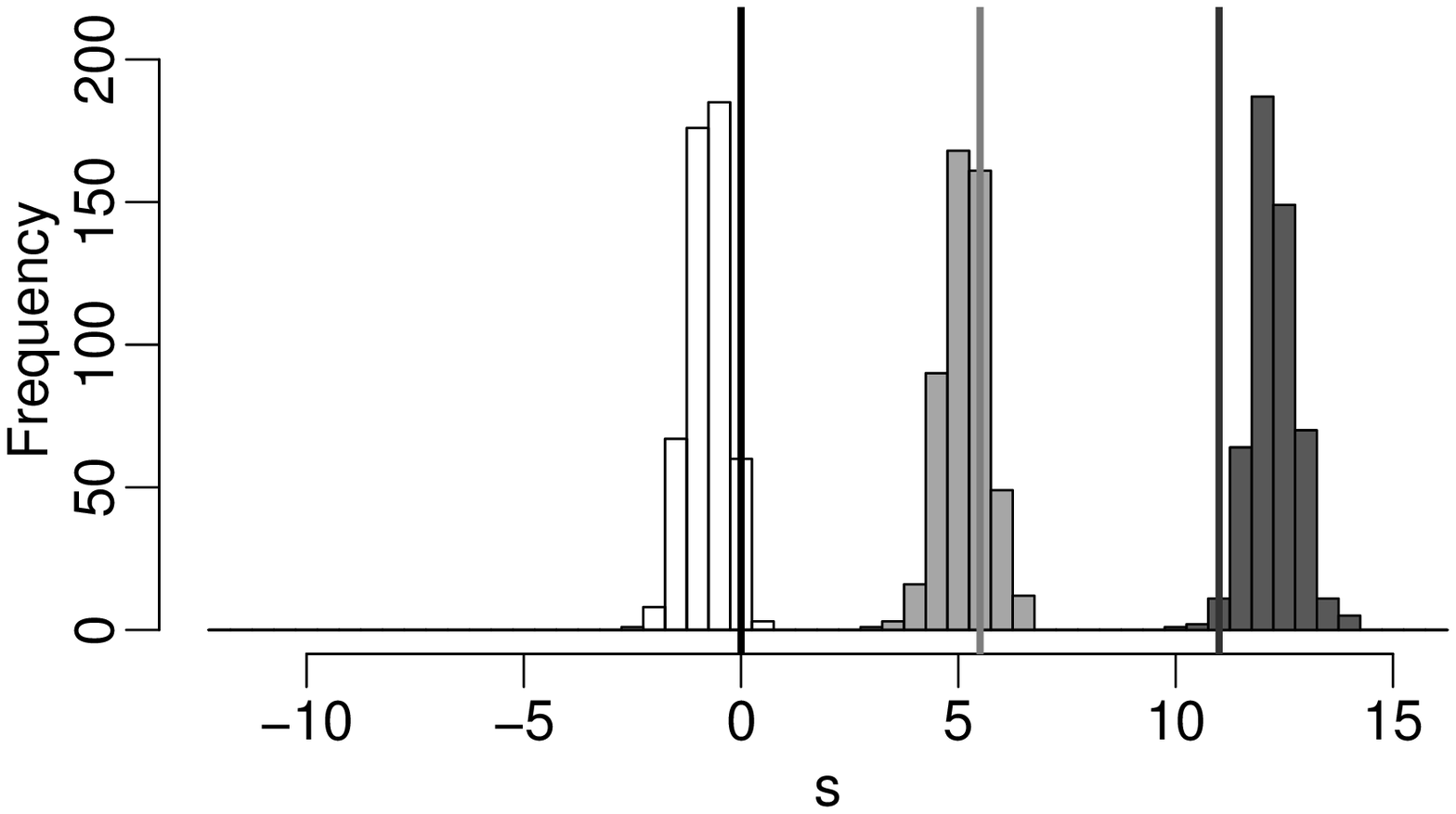}
\caption{\label{fig:simresults} Histograms showing estimates of the selection parameter $s$ from 500 simulations of data with true values of $s=0.0$ (white), $s=5.5$ (light gray), and $s=11.0$ (dark gray) when $\pi_q(\cdot)$ is chosen to reflect the distribution of initial allele frequencies used to generate the data.  The vertical lines indicate the true values of $s$.} 
\end{center}
\end{figure}

For each of the three simulated values of $s$ (0.0, 5.5, and 11.0) and each choice of $\pi_q(\cdot)$ we generated a posterior sample for each of the 500 simulated data sets.  The posterior mean of $s$ was then computed for each of these samples, yielding 500 estimates of $s$ for each simulated value of $s$ and choice of $\pi_q(\cdot)$.  Histograms of these posterior estimates when $\pi_q(\cdot)$ was chosen to approximate the initial allele frequency used to generate the data can be seen in Figure~\ref{fig:simresults}.  For all three values of $s$ the empirical distribution of the estimator covers the true value of $s$.  However, the empirical distribution of the estimator is not always centered at the true value of $s$.  Thus, this approach appears to be able to successfully distinguish among weak, strong, and moderate selection, but it may not yield an unbiased estimator of $s$. 

Additionally, a 99\% credible interval for $s$ was computed from the posterior sample of each replicate.  When $s=0.0$, 97.2\% of these intervals included the true value of $s$, whereas for $s=5.5$ and $s=11.0$ the coverage was 99.2\% and 84.4\%, respectively.  In all three cases the average length of the credible intervals was about three units.  Thus, the lower coverage for $s=11.0$ is presumably due not to narrower intervals, but rather to the fact that those intervals tend to lie above $s=11.0$.  Ultimately, our approach may lead to slightly biased estimates, but we feel that we effectively quantify the uncertainty of the estimates.

\begin{figure}[!h]
\begin{center}
\includegraphics[width=3.2in]{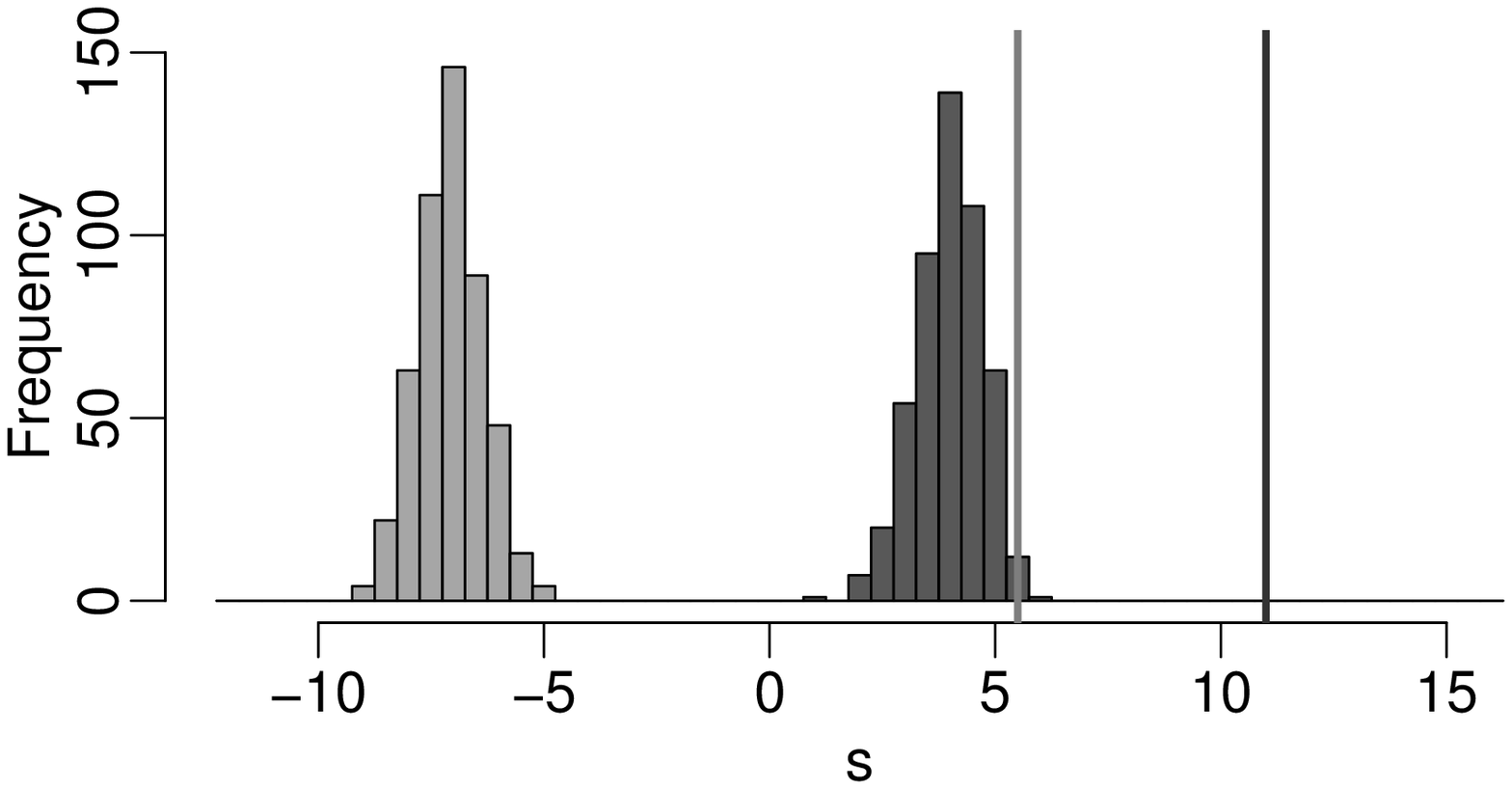}
\caption{\label{fig:unifsimresults} Histograms showing estimates of the selection parameter $s$ from 500 simulations of data with true values of $s=5.5$ (light gray) and $s=11.0$ (dark gray) when $\pi_q(\cdot)$ is chosen to be Uniform over $\mathcal{Q}_0$.  The vertical lines indicate the true values of $s$.} 
\end{center}
\end{figure}

Histograms of the posterior estimates for the cases of $s=5.5$ and $s=11.0$ when $\pi_q(\cdot)$ was chosen to be a discrete Uniform can be seen in Figure~\ref{fig:unifsimresults}.  The plot for $s=0.0$ is excluded because all of the Markov chains became trapped in a state with $s=-12.0$, which is the lower boundary of our set $\mathcal{S}$ of plausible values for $s$.  This suggests that our posterior estimate for $s$ would have been less than $-12.0$ if the algorithm had been able to search for lesser values of $s$.  Further, none of the 500 replicates for any of the simulated values of $s$ yields a single 99\% credible interval that captures the true value of $s$.  It is evident from these results that our approach relies on an accurate assumption for the distribution of the initial allele frequencies.  Assuming a discrete Uniform distribution for $q_{0,k}$ puts more mass on higher values of $q_{0,k}$ than the distribution used to generate the data did.  Consequently, we estimate values for $s$ that are too small when we assume such a distribution for $q_{0,k}$. We address the dependence of our method on an accurate choice for $\pi_q(\cdot)$ in Section~\ref{sec:discussion}.

\noindent \textbf{Example Data: Hypoxia in Flies.} An example to which this methodology can be applied is a study conducted by \citet{zhouetal2011} in which flies (\textit{Drosophila melanogaster}) were subjected to a hypoxic environment or a normal oxygen environment for $200$ generations.  Flies in the hypoxic environment were under selective pressure to adapt to the lack of oxygen in their environment.  The data consist of allele frequencies at the end of the study among approximately 200 surviving flies in each population.  These frequencies were measured at 75,999 loci along the 2L chromosome for the population under selective pressure and 71,335 loci along the 2L chromosome for the neutral population.  A subsequent analysis of these data conducted by \citet{ronenetal2013} identified a region of the 2L chromosome showing strong evidence of selection.  Within this region we have measurements at 459 loci for the population under selective pressure and measurements at 324 loci for the neutral population.  

We applied our proposed methodology to the region of the 2L chromosome of the fly data of \citet{zhouetal2011} that \citet{ronenetal2013} identified as being subject to selection.  We used the same settings as in the first set of simulations including choosing $\pi_q(\cdot)$ to be a discretized version of the distribution of allele frequencies at the end of the study for the neutral population.  Thus, we assumed that both the neutral population and the population under selective pressure began the study with a distribution of allele frequencies similar to that of the neutral population at the end of the study.  As such, we should estimate a selection parameter near zero for the neutral population.  A Markov chain was run to obtain a posterior sample of size 900 for each of the two populations of flies.  

\begin{figure}[!h]
\begin{center}
\includegraphics[width=3.2in]{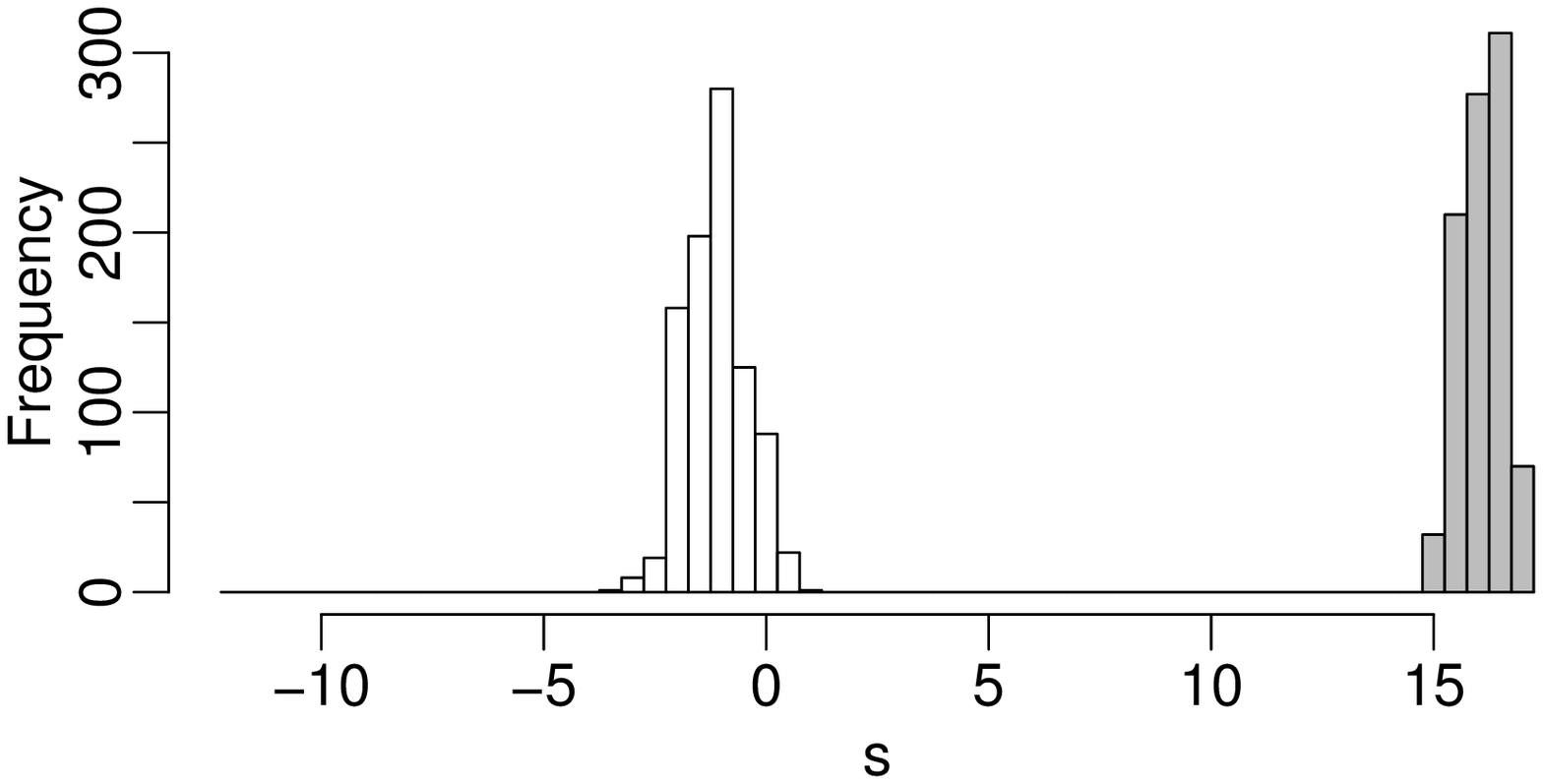}
\caption{\label{fig:flyresults} Histograms of the posterior sample of $s$ from fitting our model to the neutral fly population (white) and the fly population subjected to a hypoxic environment (gray).} 
\end{center}
\end{figure}

Figure~\ref{fig:flyresults} shows a histogram of the marginal posterior sample of $s$ for each population of flies.  These posterior samples correspond to the region of the 2L chromosome identified by \citet{ronenetal2013} as being subject to selection.  The white bars show the posterior sample for the neutral population whereas the gray bars  show the posterior sample for the population subjected to the hypoxic environment.  For the neutral population the posterior mean for $s$ is $-1.13$ and a corresponding 99\% credible interval contains zero.  Thus, as expected, there is no evidence of selection.  For the population subjected to the hypoxic environment the posterior mean for $s$ is $16.10$ and the entire sample from the posterior density of $s$ lies above zero.  This indicates strong positive selection for this population.

Finally, to determine if any other regions of the 2L chromosome display evidence of selection we applied our proposed methodolgy to sliding windows of $K=500$ loci with each window shifted 250 loci from the preceding window.  This was done for the population subjected to the hypoxic environment. Note that this procedure assumes that all loci within a sliding window have identical values for the selection coefficient, a condition that is unlikely to be strictly true.  However, we view this as a reasonable approximation given that the relatively small window size for this fairly dense SNP sample means that SNPs within a single window are likely to be linked.  In this situation, the selection coefficient estimated in our model can be viewed as an ``average'' value for the strength of selection across the loci included in the window. Thus, our method can be  used to identify regions of the genome that are likely to be under selection, but does not necessarily pinpoint individual SNPs. For each window, a Markov chain was run for 10,000 steps and every tenth step after the first 1,000 was retained to form a posterior sample of size 900.  These posterior samples were used to produce 99\% credible intervals for $s$.  

\begin{figure}[!h]
\begin{center}
\includegraphics[width=3.2in]{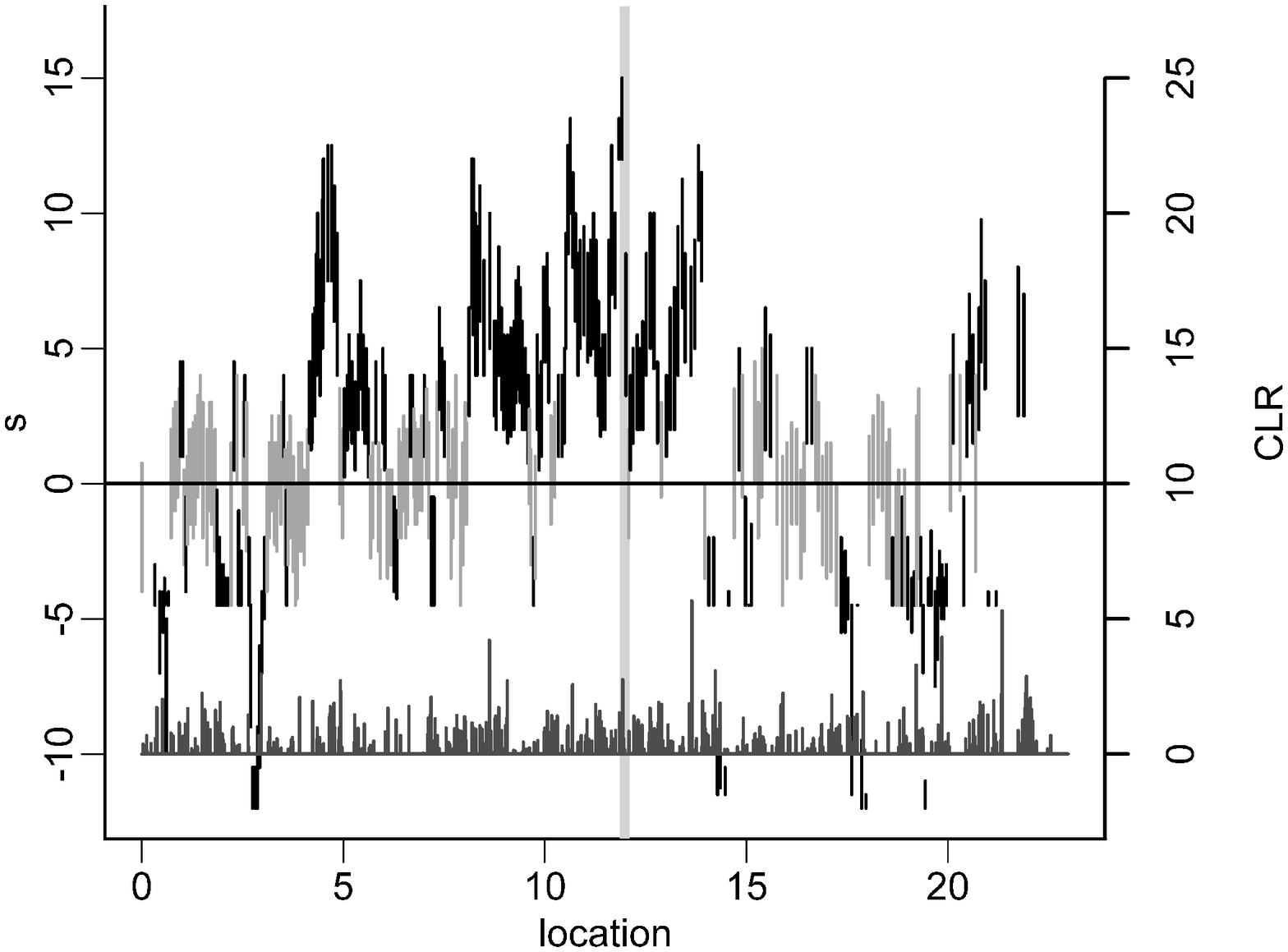}
\caption{\label{fig:windowresults}  Results of our analysis of selection for the 2L chromosome of the fly population subjected to the hypoxic environment. Location (x-axis) is displayed in units of $10^6$. The left-handed y-axis gives units for 
99\% credible intervals for $s$ for sliding windows along the chromosome.  Intervals containing zero are gray whereas those not containing zero are black.  The right-handed y-axis gives units for the composite likelihood ratio statistic computed by the SweeD software \citep{pavlidisetal2013}, shown in medium-gray at the bottom of the plot. The vertical gray bar indicates a region previously identified by \citet{ronenetal2013} as being under positive selection, which corresponds to the strongest signal of selection detected by our analysis. }
\end{center}
\end{figure}

Figure~\ref{fig:windowresults} displays 99\% credible intervals for $s$ for the sliding windows along the 2L chromosome for the population subjected to the hypoxic environment.  Intervals displayed in gray contain zero whereas those shaded black do not.  We find evidence of selection in more regions than \citet{ronenetal2013} did, including several regions under negative selection.  However, we detect the strongest signal in the same region that \citet{ronenetal2013} identified. In fact, the credible interval that lies furthest above zero falls directly in the relatively narrow region (highlighted by a gray vertical bar in Figure~\ref{fig:windowresults}) that \citet{ronenetal2013} identified using the XP-SFselect method.  To further compare our method to existing approaches, we used the SweeD software \citep{pavlidisetal2013} with grid size 30,000 to compute the composite likelihood ratio (CLR) at various points along the chromosomal region (shown in medium-gray at the bottom of Figure~\ref{fig:windowresults} with scaling given by the right-handed y-axis).  In general, the magnitude of the CLR statistic corresponds to the signal indicated by our credible intervals.

\section{Discussion}
\label{sec:discussion}
Here we propose a Bayesian method for inference of population genetics parameters under the Wright-Fisher diffusion model. In particular, we consider estimation of the selection coefficient from allele frequency data sampled at a single time point by constructing an MCMC algorithm that allows one to draw from the joint posterior distribution of the selection coefficient and the allele frequencies. We show that  when assumptions about the initial allele frequencies are accurate our method performs well in simulations as well as in the analysis of an empirical data set on hypoxia in flies. 

Existing methods for inference in similar settings have used a maximum likelihood framework and have relied on numerical approximation of the solution of the SDE corresponding to the WF diffusion.  In addition to requiring significant computation, a drawback of such methods is that they do not  provide a natural way to assess variability in the estimates. Recent work in this area has involved the development of bootstrap procedures for estimation of error variances \citep{coffmanetal2015,robinsonetal2014}, requiring significant additional computation time. An advantage of a Bayesian approach is that MCMC can be used to approximate the entire posterior distribution, providing an automatic assessment of uncertainty.  

Bayesian approaches to this problem  require an efficient method for sampling from the WF diffusion or for accurately approximating its transition density.  The important work of \citet{jenkinsspano2015} provides a method to draw exact samples from WF diffusions that incorporate features of the evolutionary process such as mutation and selection.  Their work will enable much more general developments in the application of Bayesian methods to population-level diffusion models. Our example on hypoxia in flies demonstrates the potential utility of these approaches; for example, we compute 99\% credible intervals along a chromosomal region using a sliding window approach, and recover evidence for strong positive selection in a region identified in previous studies.

Although the recent work of \citet{jenkinsspano2015} enables MCMC-based Bayesian inference in this setting, the method we have developed based on this work is also fairly computationally intensive.  This is because of the difficulty of obtaining exact draws from the general WF diffusion model.  To deal with this, we chose to use kernel density estimates of the transition density, and most of our computational effort is devoted to computing these kernel density estimates.  The computational effort needed for this step will vary depending on the true underlying parameters that generated the data. For our empirical data set for which $s \approx 16$, the method required a great deal of effort because $s$ is large compared to the neutral Wright-Fisher model, and sampling from the WF diffusion is inefficient. When the true value of $s$ is closer to zero, transition density estimates for larger values of $s$ will not be needed, and the overall computation time will be reduced.  Also, although we have exploited GPU computing in order to make this approach computationally feasible, it is possible that more efficient algorithms could be developed to speed computation further.  It may also be possible to use alternative approaches to the kernel density estimates we have used that might provide similar accuracy with reduced computational expense.

Our simulation studies clearly demonstrate that the choice of prior distribution for the initial allele frequencies across loci will have a strong effect on the posterior distribution. This is intuitive in the case of selection:  a large initial allele frequency and strong negative selection can yield the same ending allele frequency distribution as a small initial allele frequency with strong positive selection. Thus, two markedly different processes can result in very similar data, and it is not possible to disentangle the effects of these two parameters. This means that a requirement for effective use of this method is either prior knowledge of the approximate allele frequency distribution at neutrality, or a thorough exploration of the sensitivity of the results to the choice of prior distribution. Fortunately, many studies in this area do devote effort to the assessment of neutral processes in the populations under consideration and thus there will often be prior information about the distribution of initial allele frequencies available to guide choice of the prior distribution \citep[see, e.g.,][]{listephan2006, williamsonetal2005}. 

The method presented here addresses perhaps the simplest scenario one might consider in this setting: a single population and a single parameter of interest (the selection coefficient).  In practice, however, data may be collected for multiple populations or species at multiple time points, and inference for all relevant parameters (e.g., divergence times, migration rates, and selection coefficients) may be desired. While diffusion models such as this are now commonly used to model up to three populations and to infer multiple parameters simultaneously, they are generally computationally demanding and, as mentioned above, the current likelihood-based approaches require significant additional computation in order to carry out complete statistical inference that includes an assessment of uncertainty. As genome-scale data become more widely available and sampling of individuals within populations allows more precise estimates of population allele frequencies throughout the genome, methods based on allele frequencies provide a unique opportunity for efficient population-level evolutionary inference.  Methods that utilize a Bayesian framework for these problems have the potential to contribute in important ways to these developments, and our Bayesian framework provides a useful starting point for these future developments. 

\section{Acknowledgement}
This work was supported by National Science Foundation grants DMS-1209142 and DMS-1407604 (R.H.) and DMS-1106706 (L.S.K.). 

\section*{Appendix}
\label{sec:appendix}

\subsection*{Exact simulation of the neutral Wright-Fisher diffusion}

Consider a neutral WF diffusion process $Q_t$
\begin{equation}
\label{eq:wfneutral}
\left\{
\begin{array}{l}
dQ_t = \beta_0(Q_t)dt + \sqrt{Q_t (1-Q_t)} dB_t\ , \\
Q_0 = q_0, \quad 0\le t\le T
\end{array}
\right.
\end{equation}
where the drift function is $\beta_0(x) = (1/2)(\theta_1(1-x) - \theta_2 x)$, for $0\le x\le 1$ and $\theta_1, \theta_2 >0$ (note the absence of the selection parameter). The necessary details for simulating a variate $Q_T$ from the model \eqref{eq:wfneutral} for a fixed $T>0$, given that $Q_0 = q_0\in(0,1)$ are given in \citet{jenkinsspano2015} and we briefly review the main ideas. It is well-known \citep{griffiths1979,tavare1984,ethiergriffiths1993,griffithsspano2013} that, conditional on $Q_0  = q_0$, the variate $Q_T$ has a probability density function given by
\begin{equation}
\label{eq:pdfWF0}
f(x\giv q_0) = \sum_{m=0}^\infty q_m^\theta(T) \sum_{l=0}^m {\mathcal Bin}(l;m,q_0){\mathcal Beta}(x;\theta_1+l, \theta_2+m-l)
\end{equation}
where $\theta = \theta_1 + \theta_2$, and ${\mathcal Bin}(\cdot)$ and ${\mathcal Beta}(\cdot)$ are the Binomial and Beta density functions, given by
\begin{align*}
{\mathcal Bin}(l;m,q_0) &= {m\choose l} q_0^l (1-q_0)^{m-l} \\
{\mathcal Beta}(x;\theta_1, \theta_2)&\propto x^{\theta_1 - 1}(1-x)^{\theta_2 -1}\ .
\end{align*}
The mixture weights $q_m^\theta(T)$ in \eqref{eq:pdfWF0} have an alternating series expansion
\begin{equation}
\label{eq:q0}
q_m^\theta(T) = \sum_{k=m}^\infty (-1)^{k-m}a_{km}^\theta e^{-k(k+\theta-1)T/2}
\end{equation}
where
\begin{equation*}
a_{km}^\theta = \frac{(\theta+2k-1)(\theta+m)_{(k-1)}}{m!(k-m)!}
\end{equation*}
with the convention that $a_{(x)} \equiv \Gamma(a+x)/\Gamma(a)$ for $a>0$ and $x\ge-1$. Sampling from the density \eqref{eq:pdfWF0} is achieved via Agorithm \ref{alg:WF0} described below \citep[see][]{jenkinsspano2015}.

\begin{algorithm}
\caption{Exact Sampling for the neutral WF diffusion:}
\label{alg:WF0}
\begin{algorithmic}[1]
\State Draw $\sM$ from the discrete distribution $\{q_m^\theta(T), m=0,1,\dots\}$;
\State Draw $\sL \sim {\mathcal Bin}(\sM, q_0)$;
\State Draw $\sQ \sim {\mathcal Beta}(\theta_1+\sL, \theta_2+\sM-\sL)$;
\State \textbf{Return} $\sQ$.
\end{algorithmic}
\end{algorithm}

The computational burden in Algorithm \ref{alg:WF0} is in Step 1, since the weights $q_m^\theta(T)$ do not have a closed form expression and \eqref{eq:q0} suggests an infinite amount of computation. However, given the alternating series representation in \eqref{eq:q0}, one can use ideas described in Chapter 5 of  \citet{devroye1986} to devise an exact procedure for simulating the variate $\sM$, which only requires finite computing time.   Details of such a procedure are provided in Section~3.2 of \citet{jenkinsspano2015}.   

\subsection*{Exact simulation of general WF-diffusion models}
As we mention above, our approach relies on being able to obtain exact draws from a general Wright-Fisher diffusion \eqref{eq:WFgeneric}. We briefly review the general setup for sampling an SDE. Assume that $\{X_t,\ 0\le t\le T\}$ is a stochastic process described via 
\begin{equation}
\label{eq:sde}
dX_t = \beta(X_t)dt + \sigma(X_t)dB_t, \quad\quad X(0) = x_0\ , \quad 0\le t\le T
\end{equation}
where the functions $\beta(\cdot)$  and $\sigma(\cdot)$ are assumed to be smooth enough such that \eqref{eq:sde} has a unique weak solution. Such conditions can be found in \citet{karatzas2012brownian}, for example. Our goal is to simulate exact draws from the distribution of $X_t$, for some fixed $t>0$, conditional on $X_0=x_0$. In general, the transition density for the process $X_t$ is not available in closed form, not even in an infinite series representation as is the case with the neutral Wright-Fisher diffusion. Recently, in a series of papers \citep{EA1,EA2,EA3}, the authors have developed a rejection sampling approach, which yields an exact (distribution-wise) skeleton of the full path $(X_t, 0\le t\le T)$ for a very large class of diffusions. Briefly, let $\Omega = {\mathcal C}([0,T])$ and let $\omega$ be a typical element of $\Omega$. Let $\QQ$ be the probability measure induced by $\{X_t,\ 0\le t\le T\}$ onto $\Omega$ and $\ZZ$ be another probability measure on $\Omega$ which is user-selected in an appropriate way. Under regularity conditions \citep[see][]{EA1} one can use the Girsanov formula to derive and expression for the Radon-Nikodym derivative
$$
\frac{d\QQ}{d\ZZ}(\omega) \propto {\mathsf G}(\omega)
$$
where, in principle, one can arrange that ${\mathsf G}(\cdot) \in (0,1)$. Thus, a rejection sampling strategy will be appropriate:

\begin{algorithm}
\caption{Exact sampling for general diffusions}
\label{alg:rejection}
\begin{algorithmic}[1]
\State Sample $\omega\sim \ZZ$;
\State Accept $\omega$ as a draw from $\QQ$ with probability $\sG(\omega)\in (0,1)$.
\end{algorithmic}
\end{algorithm}

The proposal distribution $\ZZ$ has to be selected in a way that Step 1 of Algorithm \ref{alg:rejection} can be done efficiently, i.e. a biased Brownian Bridge measure. We refer the reader to \citet{EA1} for a detailed description of all of the conditions and the general setup. 

\vspace{0.5cm}
\noindent
\textit{General Wright-Fisher diffusions}. If the diffusion coefficient in \eqref{eq:sde} takes the form
$$
\sigma(x) = \sqrt{x(1-x)},
$$
then the SDE \eqref{eq:sde} is a general WF diffusion. As \citet{jenkinsspano2015} suggest, in this case, it is efficient to select the proposal measure $\ZZ$ to be the law induced by the neutral WF diffusion \eqref{eq:wfneutral}.

\bibliographystyle{apalike}
\bibliography{references}

\end{document}